\documentclass[aps,pra,twocolumn,superscriptaddress]{revtex4-1}

\usepackage{mathptmx}
\usepackage{subfigure}
\usepackage{dcolumn}
\usepackage{amsmath,amssymb}
\usepackage{bm}
\usepackage{color}
\usepackage{latexsym}
\usepackage[english]{babel}
\usepackage{times}

\usepackage{psfrag,graphicx} 
\usepackage{epsf}  
\usepackage{amsfonts}
\usepackage{natbib}
\usepackage{epstopdf}
\usepackage{appendix}
\usepackage[colorlinks=true,citecolor=blue,linkcolor=blue]{hyperref}

\DeclareMathOperator{\arccosh}{arccosh}

\begin{document}

\title{Optical lattices with large scattering length: Using few-body physics to simulate an electron-phonon system}

\author{Zhihao Lan}
\email{z.lan@soton.ac.uk, lanzhihao7@gmail.com}
\affiliation{Mathematical Sciences, University of Southampton, Highfield, Southampton, SO17 1BJ, United Kingdom}

\author{Carlos Lobo}
\affiliation{Mathematical Sciences, University of Southampton, Highfield, Southampton, SO17 1BJ, United Kingdom}

\date{\today}
\newcommand{\ad}{\hat{a}^\dagger}
\newcommand{\ha}{\hat{a}}
\newcommand{\hn}{\hat{n}}
\newcommand{\bE}{{\bf E}}
\newcommand{\bB}{{\bf B}}
\newcommand{\bJ}{{\bf J}}
\newcommand{\bA}{{\bf A}}
\newcommand{\bS}{{\bf S}}
\newcommand{\bR}{{\bf R}}
\newcommand{\br}{{\bf r}}
\newcommand{\bk}{{\bf k}}
\newcommand{\hbx}{\hat{\bf x}}
\newcommand{\hby}{\hat{\bf y}}
\newcommand{\hbz}{\hat{\bf z}}
\newcommand{\hbr}{\hat{\bf r}}
\newcommand{\hbn}{\hat{\bf n}}
\newcommand{\hbk}{\hat{\bf k}}
\newcommand{\hbS}{\hat{\bf S}}
\newcommand{\hS}{\hat{S}}
\newcommand{\hbsigma}{\hat{\bf \sigma}}
\newcommand{\bp}{{\bf p}}
\newcommand{\brp}{{\bf r}^\prime}
\newcommand{\bF}{{\bf F}}
\newcommand{\pe}{\frac{1}{4 \pi \epsilon_0}}
\newcommand{\vs}{\vspace{0.25cm}}
\newcommand{\ee}{\end{equation}}
\newcommand{\be}{\begin{equation}}
\newcommand{\en}{\mathcal{E}}
\begin{abstract}
We propose to go beyond the usual Hubbard model description of atoms in optical lattices and show how few-body physics can be used to simulate many-body phenomena, e.g., an electron-phonon system. 
We take one atomic species to be trapped in a deep optical lattice at full filling and another to be untrapped spin-polarized fermions (which do not see the optical lattice) but has an s-wave contact interaction with the first species. For large positive scattering length on the order of lattice spacing, the usual two-body bound (dimer) states overlap forming giant orbitals extending over the entire lattice, which can  be viewed as an ``electronic" band for the untrapped species while the trapped atoms become the ``ions" with their own on-site dynamics,  thereby simulating an electron-phonon system with renormalization of the phonon frequencies and Peierls transitions. This setup requires large scattering lengths but minimises losses, does not need higher bands and adds new degrees of freedom which cannot easily be described in terms of lattice variables, thus opening up intriguing possibilities to explore interesting physics at the interface between few-body and many-body systems.

\end{abstract}

\maketitle 

\section{introduction}

Experimental studies of optical lattices (OLs) do not usually require the existence of Feshbach resonances since interactions between atoms in the lattice must be weak (the scattering length is much smaller than the lattice spacing so as not to occupy higher bands \cite{ETH}). Here we will study an OL where the presence of a Feshbach resonance is required since we will need to tune one of the scattering lengths to be much larger than the lattice spacing $a \gtrsim d$ while keeping the single band approach. We confine one atomic species to the OL while a second species interacts with the first one but is untrapped. The second species forms {\em bound states} with the first and is trapped only by interactions \cite{Petrov, Antezza} (Fig. \ref{f1}). It requires achieving $a \sim d$ for which we need a Feshbach resonance with good magnetic field control \cite{rudi, randi}. This setup brings new tools to the many-body problem allowing for new non-lattice degrees of freedom which mediate interactions between trapped atoms, going beyond the conventional Hubbard model. It can also be seen as bringing few-body physics into a new setting by giving molecules enhanced stability due to the lattice since it strongly suppresses three-body losses. Efimov physics and other few-body phenomena have up to now only been studied by measuring losses, and no stable trimer has ever been trapped due to three-body recombination. The stability against losses created by spatially separating the different species using external potentials was already known from previous studies in few-body systems \cite{mixed_dimension, nishida, wei_zhang} where confinement to 1D tubes or 2D planes is used. Here we instead explore the interface between few-body and many-body physics.

As an example we implement an electron-phonon quantum simulator where the trapped atoms are the ``ions" and the untrapped ones the ``electrons" which are spin polarized fermions to minimise losses. The trapped atoms are in a deep OL such that their wave functions do not overlap and their statistics are unimportant. Without the ``electrons" they oscillate at the lattice onsite frequency (flat phonon dispersion); yet, when they interact with each other via the ``electrons" (Fig. \ref{f1}b), they exhibit collective phonon oscillations with frequency dispersion. Depending on the filling factor of the untrapped species to the ``electronic" band, other phenomenology, such as the Peierls instability (at half-filling) leading to a dimerized phase as well as polaron physics (when only one untrapped atom is present), is also possible in this simulator. 

A traditional way of simulating electron-phonon systems with cold atoms is to trap the species representing electrons in an OL and place it in contact with the Bose-condensate of another species that provides the  phonons \cite{Jaksch_free1, Jaksch_free2, Jaksch_free3, tilted_lattice,Hofstetter_lattice, Jaksch_lattice1, Jaksch_lattice2}. In our case the trapped species represents the ions providing the phonons and the untrapped species represents the electrons. A few important differences arise. For example, in the traditional case, an acoustic phonon branch exists, whereas here the trapping of the ``ions" leads to an optical branch. Furthermore, the interaction between the trapped atoms mediated by the background Bogoliubov modes of the condensate  is always attractive regardless of the sign of the trapped atom-condensate interaction \cite{Jaksch_free1, Jaksch_free2, Jaksch_free3, Hofstetter_lattice}. In our case, a prominent feature is that the mediated interaction between trapped atoms can be tuned from repulsive to attractive by changing the filling of the untrapped atoms to the ``electronic" band conveniently. Also, here the lattice for the heavy atoms allows us to study solid state issues such as the influence of vacancies on the phonon spectrum. Beyond this of course, the underlying physics behind the simulators is quite different since this work uses the tools of strongly interacting few-body systems. Other theoretical proposals for electron-phonon quantum simulator have used, for example, highly excited Rydberg states of cold fermionic atoms in a bilayer lattice \cite{hague}, or hybrid systems composed of a crystal of trapped ions coupled to a gas of ultracold fermions \cite{gerritsma}. To the best of our knowledge none of the simulators proposed so far has been implemented experimentally.

As we mentioned above, the idea of trapping one species while allowing the other to be free has been explored in a number of contexts previously. In \cite{Petrov}, heavy atoms were confined to a lattice and each of them formed a dimer with an unconfined light one. The original idea was merely to trap the heavy atoms to a 2D plane so that, at sufficiently high density, the dimer-dimer repulsion would lead to a molecular solid. Unfortunately the dimer mass was too low and it was necessary to add an external optical lattice potential in the plane to increase the heavy atom in-plane effective mass, allowing for crystallisation. In order to avoid commensuration problems, in this scheme, the lattice filling must be very low and tunnelling must be large, unlike the present case where we will require unit filling and no tunnelling (deep lattice). Note also that in  \cite{Petrov}, the number of heavy atoms must be equal to that of light atoms since only heavy-light dimers exist. Here however, the stability is given by the deep lattice (and not by dimer-dimer repulsion) which will allow us to consider situations where the number of light atoms can be a fraction of that of heavy ones. For us therefore, the optical lattice is a real physical object in the calculation, not just a shift in the band mass. In other works, the lattice is sufficiently deep that tunnelling was not allowed although the focus was on bound states in the disordered case without onsite dynamics \cite{Antezza}, while in \cite{Massignan}, the two species scatter off each other and do not form bound states. Apart from these many-body studies, there are some few-body ones as discussed above \cite{mixed_dimension, nishida, wei_zhang} where the stability of the deep lattice is explored to study trimers exhibiting Efimov physics.

\begin{figure}[!hbp]
\centering
\includegraphics[width=1.0\columnwidth]{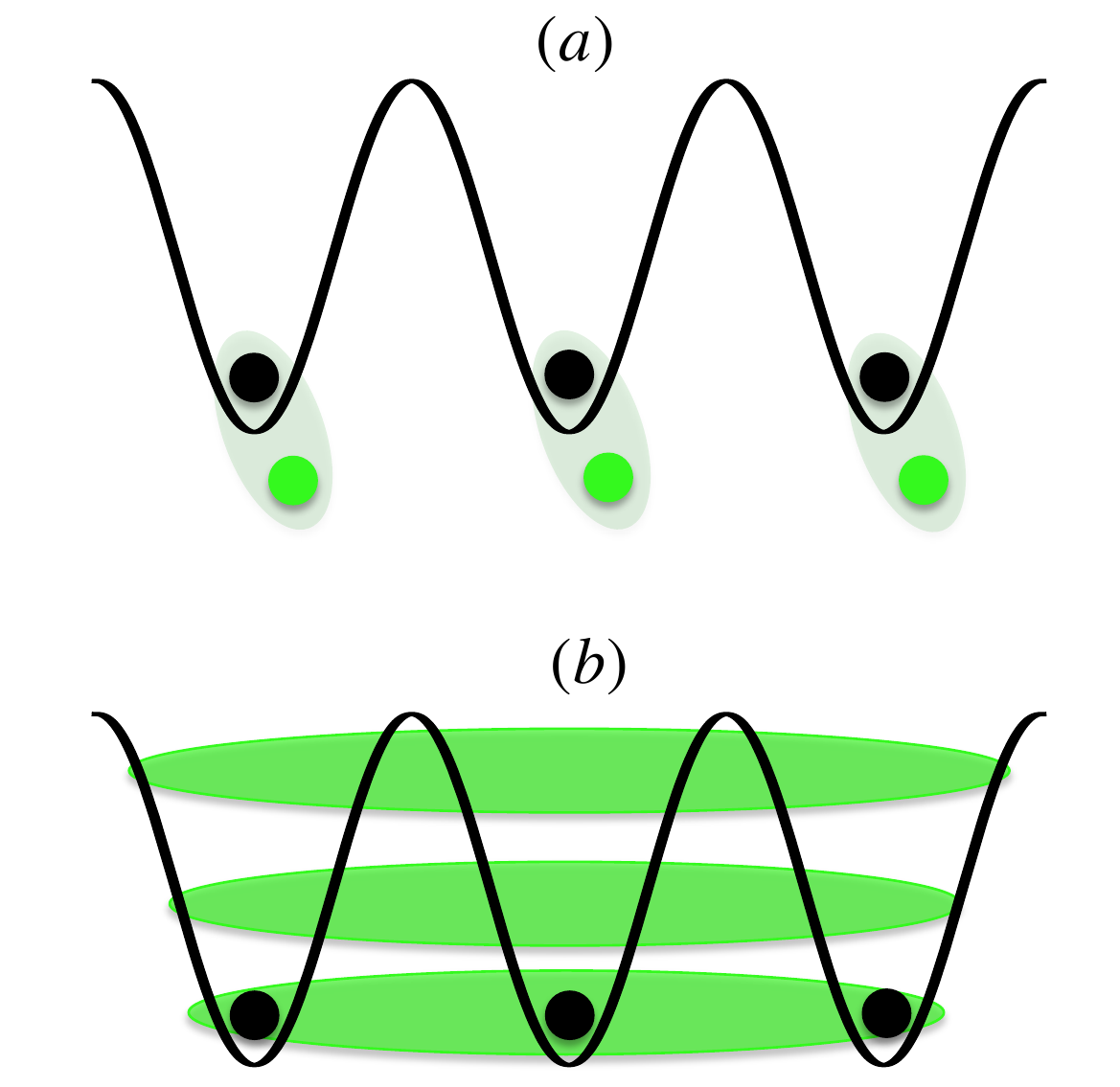}
\caption{ (Color online) (a) Dimers in an optical lattice: the heavy atoms (black dots) are trapped in a deep lattice and interact with untrapped light atoms via an s-wave contact interaction such that they form Feshbach molecules. (b) When the scattering length between the two species is on the order of lattice spacing, the usual two-body bound states overlap forming giant orbitals (or Bloch states - green shaded areas) extending over the entire lattice which form a band. The mediated interaction between the trapped heavy atoms crucially depends on the filling of the light atoms to the band.}
\label{f1}
\end{figure}

The remaining parts of the paper are organized as follows.  In section \ref{s2}, we introduce the theoretical framework of our study, which is based on the Born-Oppenheimer approximation (BOA). In section \ref{s3}, the energy band structure for the untrapped species in the BCS (Bardeen-Cooper-Schrieffer)-BEC (Bose-Einstein condensate) crossover is presented in the limit where the trapping of the OL is sufficiently strong that the trapped atoms can be considered as fixed scatterers. In section \ref{s4}, we relax this condition and allow for onsite motion of the trapped atoms and study the resulting phonon dynamics. We then show that this dynamics can be mapped on to the quantum transverse Ising model and a method to calculate the interactions of the heavy atoms at full-filling and half-filling is given. Two interesting phases of this spin model are discussed: the ferromagnetic (FM) and anti-ferromagnetic (AFM) phases which correspond to the normal and dimerized lattice of the heavy atoms. The experimental issues of this work are discussed in section \ref{s5} and we conclude in section \ref{s6}.

\section{The model}
\label{s2}
The setup we considered is shown in Fig. \ref{f1} where one species is trapped by an OL while the other is untrapped but has an s-wave contact interaction with the first.  We assume that the trapped atoms are heavier than the untrapped ones ($M  \gg  m$) in order to use the BOA \cite{BOA}, but this is not an experimental requirement: for any mass ratio we expect on general grounds that the electron-phonon physics simulated will remain qualitatively the same even if the BOA is no longer applicable. The validity of the BOA will be discussed below.  In the BOA, we start by considering the heavy atoms as fixed and solving the Schr\"odinger equation for the motion of the light atoms. The wave function of $N_l$ light and $N_h$ heavy atoms is written in the form $\Psi$=$\psi\left( \{ \bR \} \right) \chi \left( \{ \bR \}, \{ \br \} \right) $, where $\{\bR\},\{\br \}$ are the sets of coordinates of the heavy and light atoms. $\chi$ is a Slater determinant of $N_l$ light atom states (note we will work only with bound states in this paper so that these are of the decaying exponential type)
\begin{gather}
\phi(\br)  =  \sum_{j=1}^{N_h}c_j \frac{{\rm e} ^{-\kappa \| \bR_j - \br \|}}{\| \bR_j -\br \|}. 
\label{wf}
\end{gather}
We see that the light atom states are simply a sum of 3D Green's functions located at $\bR_j $ and 
\be 
\epsilon(\kappa)=-\frac{\hbar^2 \kappa^2}{2m} < 0
\ee
 is the energy of each light atom orbital. In the BOA, the sum of the eigenenergies of the light atoms provides an effective interaction between the heavy atoms so that the total heavy atom interaction energy is $\sum_{i=1}^{N_l} \epsilon(\kappa_i)$. The interaction between heavy atoms and light atoms is taken into account by using the  Bethe-Peierls boundary conditions \cite{PetrovReview}
\begin{gather}
\phi(\br\rightarrow \bR_j)  \propto \frac{1}{\| \bR_j -\br \|} -\frac{1}{a}
\end{gather}
 where $a$ is the scattering length between the two species and we will assume that it can be tuned by a magnetic Feshbach resonance \cite {Feshbach_resonance}. Applying the above boundary conditions to the wave function (\ref{wf}) leads to: 
\be
\left( \kappa-\frac{1}{a} \right) c_i=\sum_{j( \neq i)=1 }^{N_h} \frac{{\rm e}^{-\kappa \| \bR_j - \bR_i \|} }{\| \bR_j - \bR_i \|} c_j \label{eigenvalue}
\ee

\section{ light atom energy band in the BEC-BCS crossover}
\label{s3}

\begin{figure}[!hbp]
\centering
\includegraphics[width=1.0\columnwidth]{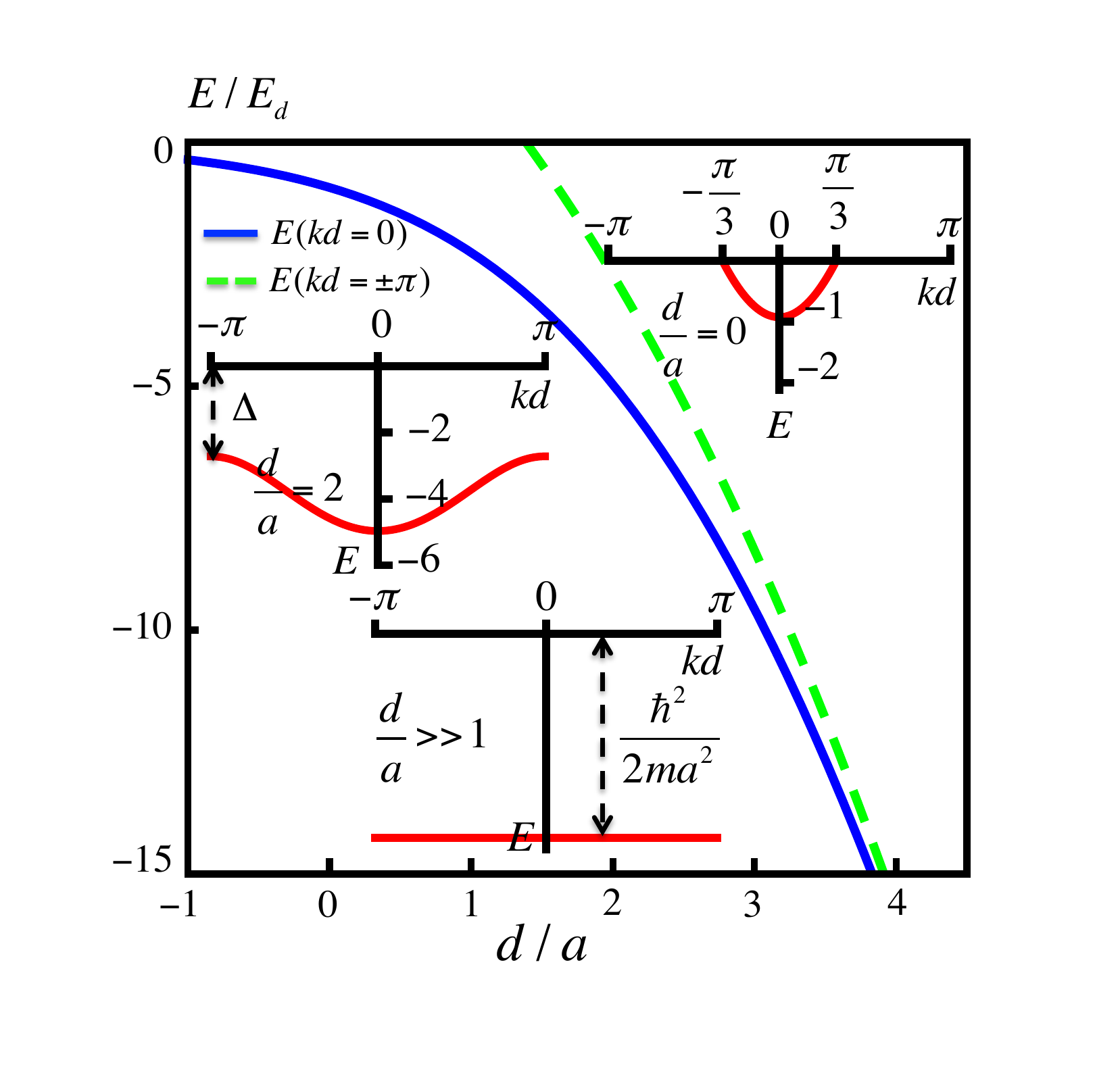}
\caption{(Color online) Energy band structure for the light atoms in the BEC-BCS crossover.  Band structure as a function of $d/a$, where the dashed curve shows the evolution of the top of the band while the solid curve shows the bottom of the band. The three inserts show the band structure at three different regimes: $d/a \gg 1$ (dimer regime); $d/a =2$ (insulator regime); $d/a =0$ (unitary regime). }
\label{electron}
\end{figure}

We first calculate the energy band structure for an ordinary OL, where the  trapping potential of the heavy atoms is a 1D lattice tightly confined radially:
\be
V=\frac{1}{2} M \omega_\perp^2(x^2+y^2) + V_0 \cos^2(\pi z/d)
\ee
so that the resulting potential is harmonic on each lattice site with $\omega_\perp  \gg  \omega_0  \equiv  \sqrt{4 E_R V_0}/\hbar$ where $\omega_0$ is the onsite lattice oscillation frequency and the recoil energy is $E_R  \equiv  \pi^2 E_d m/M $ where $E_d  \equiv  \hbar^2/2m d^2$ \cite{note1}. Therefore the radial motion of the heavy atoms can be considered to be frozen and the extent of their wave function along that direction ($\sim  \sqrt{\hbar/M\omega_\perp}$) is very small compared to its transverse size \cite{note2}, so that in the rest of the paper we will consider the wave function on each site to be essentially 1D. This system implements a Kronig-Penney-type model discussed in \cite{Kronig_Penney} since it is equivalent to a linear array of 3D $\delta$ function potentials for the light atoms.

To reduce losses we minimize the overlap of the heavy atom wave functions \cite{note3} keeping $V_0$ large so that their width along $z$ is  much smaller than $d$ (e.g., for $V_0= 25 E_R$, $\sqrt{\hbar/M\omega_0}=0.14d$). So, in calculating $\phi$, we assume that the heavy atoms are localised at the lattice site minima $\bR_j=j d \hbz$ making the potential periodic and assuming periodic boundary conditions. According to Bloch's theorem, $c_j=\exp(i k jd)$ with $\exp(i k d N_h)=1$. Replacing the expressions of $\bR_j$ and $c_j$ in (\ref{eigenvalue}), taking the $j=0$ site to be at the centre of the chain (to have both positive and negative positions) and taking $N_h \rightarrow \infty$, we get
\begin{gather}
\left( \kappa-\frac{1}{a} \right) =\sum_{j=1}^{\infty} \frac{{\rm e}^{-\kappa j d } }{jd} e^{ikdj}+\frac{{\rm e}^{-\kappa j d } }{jd} e^{-ikdj}.
\end{gather}

Using $\log(1-x)=-\sum_{n=1}^{\infty}\frac{x^n}{n}$, it is straightforward to solve $\kappa(k)$ from the above equation and  the resulting expression of the ``electron" band $E(k)=-\hbar^2 \kappa^2(k)/2m $ is very different from the single particle dispersion of a noninteracting Hamiltonian with nearest neighbour hopping,
\begin{equation}
E(k)=-E_d\arccosh^2 \left(\frac{  {\rm e}^{d/a} }{2}+\cos(kd) \right).
\label{band}
\end{equation}
\noindent

The evolution of the band structure in the BEC-BCS croosover has the following properties. In the full-filled deep ``BEC limit" ($a \rightarrow 0^+$) the atoms form $N_l$=$N_h$ tightly bound dimers of energy $-\hbar^2/2ma^2$ \cite{note4}. The dispersion is flat (the bandwidth is zero) since the dimers hardly overlap with each other (Fig. \ref{f1}a). However, as $a$ increases from zero, the bandwidth becomes non-zero, with both the bottom ($k=0$) and the top of the band ($k=\pm\pi/d)$ increasing (Fig. \ref{electron} ). The band gap to the continuum
\be
E_{\rm gap}=E_d \arccosh^2(  {\rm e}^{d/a}/2-1 ) \label{gap}
\ee
gradually decreases until it disappears at $d/a=\ln 4  \simeq  1.39$.

To study the validity of the BOA we need to estimate the importance of the nonadiabatic terms. For a full-filled band, these terms correspond to the transfer of light atoms into the continuum due to lattice deformations. However, if $E_{\rm gap}   \gtrsim  0$ then these transfers are forbidden by energy conservation. To show this, we note that the creation of a localized hole leads to an attraction between heavy atoms $\sim  E_d e^{-d/a}$. But, for $V_0=25 E_R$ and $M/m \simeq  10$, the harmonic potential of the lattice ($\sim  10 E_R \sim  10 E_d$) is much greater, making any lattice deformation very small so that the true energy gap is approximately equal to the band gap except perhaps for very small $E_{\rm gap}$.

At unitarity the BOA tells us that the gap has closed since the band now has a Fermi surface at $k=\pm \pi/3d$ and that the bandwidth is $0.93 E_d$. However, the BOA can no longer be trusted here due to the importance of the nonadiabatic terms \cite{note5}. Nevertheless it still gives us an indication that about one third of the light atoms remain bound while the other two thirds have been lost to the continuum. For any value of $d/a$ there is still a fraction of bound atoms although it becomes very small on the ``BCS" side since the bottom of the band gradually approaches zero as $d/a  \rightarrow-\infty$ \cite{note6}.

We will focus on the regime where the scattering length is on the order of the lattice spacing, i.e., $a \sim d$, since it allows for a dispersive band which will mediate the interaction between the heavy atoms while keeping a gap which is important for the stability of the system.

This novel way to implement an effective lattice potential for the light atoms via their interaction with the heavy atoms has interesting features. For example, it allows us to create a lattice for the light species using lasers which only trap the heavy atoms and as mentioned above is a perfect implementation of the Kronig-Penney (KP) model. Also, at finite temperatures, atoms with large kinetic energy have a higher probability of escaping to the continuum which could provide a natural evaporative cooling process assuming that there is a thermalization mechanism for the remaining ones, e.g., via collisions with the heavy atoms. Finally, at very low temperatures and for small enough gaps, light atoms can in principle tunnel out of the band which might lead to interesting analogies with tunnelling problems in solid state physics.

Without the OL, losses are due to either formation of few-body states of size $a$ or relaxation to deep bound dimer states. Since we are in fact working already with Feshbach bound states, losses of the first kind can only come from excitation of light atoms into the continuum due to collisions with phonons, a process which might represent evaporative cooling or dimerisation of the lattice and depends strongly on the size of the gap relative to the electron-phonon coupling strength. Relaxation to deep bound states however can always occur and in our case comes from the interaction of two light and one heavy atoms (heavy-heavy-light interactions are forbidden due to the lattice). These have been investigated in \cite{Petrov}  with the conclusion that the rate of formation was of the order of $E_d (R_e/a)^4 \exp(-2 d/a)/\hbar$ where $R_e$ is the range of the interatomic potential. For the gapped case $d/a$=$2$, this relaxation time was found to be $\gtrsim  10$s for a $^6$Li-$^{40}$K mixture.

\section{phonon dynamics of the heavy atoms}
\label{s4}
Turning to the phonons, the oscillation frequency of the heavy atoms in the lattice is changed from $\omega_0$ due to the interaction with the light atoms. To estimate the magnitude of this shift we consider $d/a=2$ where the range of the mediated heavy atom interaction is small and so we can neglect the interactions beyond nearest neighbours. In this case the interaction between two heavy atoms at a distance $R$ for full-filling is, apart from a constant term \cite{Petrov},
\be
U(R)=2 \frac{\hbar^2}{ma^2} {\rm e}^{-2 R/a} (R/a)^{-1} \left(1- \frac{(R/a)^{-1}}{2} \right). \label{Petrov}
\ee
We assume that the two neighbours of a particular heavy atom are at their equilibrium positions and calculate the frequency of oscillation of the heavy atom around equilibrium:
\be
\omega=\sqrt{\omega_0^2+\frac{2 U'' (R=2a)}{M}} \simeq \omega_0 \left(1+ \frac{U'' (R=2a)}{M \omega_0^2} \right).
\ee
This is approximately equivalent to estimating the square of the frequency shift of the $\pi$ phonon, i.e., $(\omega^2(q$=$\pi/d)-\omega_0^2)/\omega_0^2$ which is the square of the ratio of the phonon bandwidth to $\omega_0$ and is a dimensionless measure of the ``electron"-phonon coupling strength. We find that, for $V_0=25 E_R$ and therefore $\hbar \omega_0=10 E_R$, the shift is $\simeq 0.001 M/m$, i.e., in practice there will be no appreciable effect. We might try to increase the effect by reducing $\omega_0 \equiv  \sqrt{4 V_0 E_R}/\hbar$.
However, as pointed out above, $V_0$ must be large in order to minimize the overlap of wave functions of adjacent heavy atoms to reduce losses and, as we see from Fig. \ref{electron}, $d/a  \sim  O(1)$ to keep the gap open for the BOA to be valid. Therefore the band energy remains around $\sim  E_R$. This means that $\hbar \omega_0  \gg  E_R$, leading to a small shift in phonon frequencies.

\begin{figure}[!hbp]
\centering
\includegraphics[width=1.0\columnwidth]{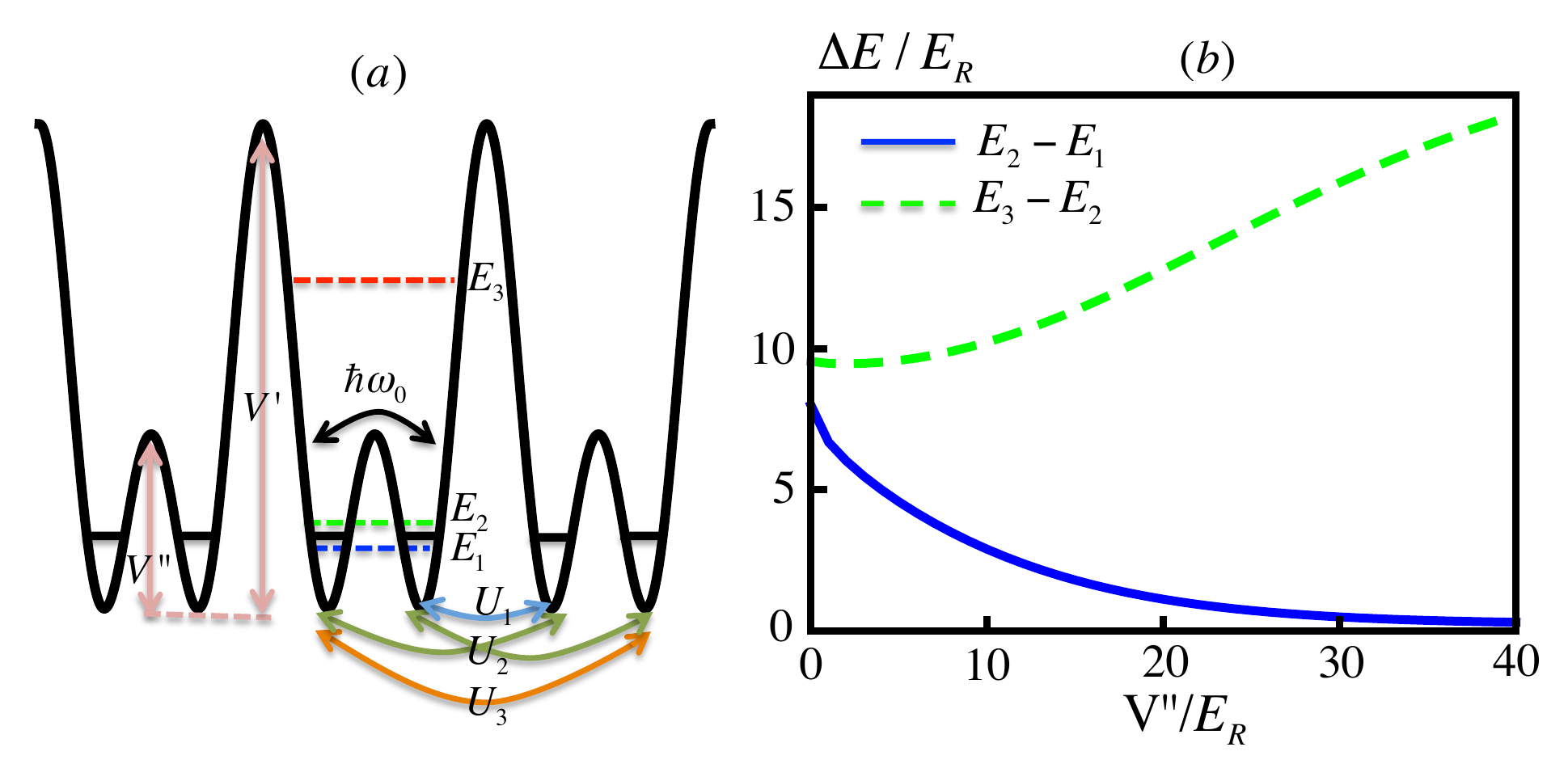}
\caption{ (Color online) (a) Superlattice scheme for phonon tuning, where $V'$ and $V''$ are the interwell and intrawell potential heights; $E_1$, $E_2$ and $E_3$, the energy levels of the double-well; $\omega_0$ the tunnelling frequency between the left and right wells; $U_1$, $U_2$ and $U_3$, the nearest-neighbour interactions of the heavy atoms in the double wells. (b) Level spacing in the double-well as a function of $V''$ for fixed $V'=40 E_R$. When $V''$ is large, $E_3-E_2  \gg  E_2-E_1  \sim  E_R$ which justifies our two-state model. }
\label{phonon}
\end{figure}

To overcome this difficulty we propose to use a superlattice where a single heavy atom is confined in each double-well:
\be
V=\frac{1}{2} M \omega_\perp^2(x^2+y^2) + V_0 \cos^2(\pi z/d) + V_1 \cos^2(2\pi z/d), \label{potential}
\ee
which describes a lattice of double-wells with intrawell tunnelling controlled by $V'' = V_1-V_0/2+V_0^2/16V_1$ and interwell tunnelling controlled by $V'=V_1+V_0/2+V_0^2/16V_1$ (Fig. \ref{phonon}a). Because of the extra parameter, we can: keep $d/a  \sim  O(1)$ and large $V'$ (so that there is no tunnelling between double-wells), and tune the intrawell tunnelling to $\sim  E_R$ \cite{Bloch}. We can further restrict ourselves to the two lowest energy states $E_{1,2}$ of the double-well since, for example, with $V''=20 E_R$ and $V'=40 E_R$, $E_3-E_2 \simeq  14 E_R  \gg  E_2-E_1 \simeq  1 E_R$ (Fig. \ref{phonon}b). The oscillation of the heavy atom in the double-well replaces the onsite oscillation in the simple OL studied above so that the level splitting corresponds to $\hbar \omega_0$. The superlattice also has the advantage of allowing for a significant displacement of the heavy atoms during the oscillation or in the charge density wave (CDW). This increases the ``electron"-phonon coupling strength and also could allow the CDW to be detected via light scattering due to the appearance of a secondary peak corresponding to its periodicity. With the parameters above, the maximum atomic displacement $\delta \langle z \rangle$ is $\simeq  0.2 d$, a significant fraction of the lattice period.

\subsection{Effective models for the heavy atoms}

We first write down the extended Hubbard model for the heavy atoms in the double-well lattice with periodic boundary conditions and within the BOA. Since the two lowest energy states $E_{1,2}$ of the double-well are separated from $E_3$ by more than $10 E_R$, we can restrict ourselves to two states per double-well and map the model to a quantum transverse Ising model. The left/right basis wave functions for site $j$, $\xi_{j,R/L}$ (the sum and difference of the two lowest eigenstates) are well localized on each side of the double-well since the tunnelling is small. As we will see later, for the full- and half-filled cases, we can keep only nearest-neighbour interactions and for one heavy atom in each double-well, we get the following extended Hubbard model for the heavy atoms in the BOA:
\begin{eqnarray}
\hat{H}=\sum_{i=1}^{N_h} &&-\hbar \omega_0 \left( \ad_{i,L} \ha_{i,R} + \ad_{i,R} \ha_{i,L} \right) +U_1 \hn_{i,R} \hn_{i+1,L} \nonumber \\
&&+U_2 \left( \hn_{i,R} \hn_{i+1,R} + \hn_{i,L} \hn_{i+1,L}\right) + U_3 \hn_{i,L} \hn_{i+1,R}
\end{eqnarray}
where $\hat{a}^{\dagger}_{i,L/R}$ ($\hat{a}_{i,L/R}$) creates (destroys) a heavy atom in the left/right well at site $i$. Similarly, $\hn_{i,L/R} \equiv \hat{a}^{\dagger}_{i,L/R} \hat{a}_{i,L/R}$ is the number operator and $\omega_0$ is the oscillation frequency in the double-well. The $U_{1,2,3}$ are given in terms of the interaction $U(R) $ (discussed in the next section) by
\begin{eqnarray}
U_1&=&\int d^3R_1 d^3R_2 |\xi_{1,R}(R_1)|^2 U(R_1-R_2) |\xi_{2,L}(R_2)|^2 \label{U1}, \\
U_2&=&\int d^3R_1 d^3R_2 |\xi_{1,L}(R_1)|^2 U(R_1-R_2) |\xi_{2,L}(R_2)|^2 \nonumber \\&=&\int d^3R_1 d^3R_2 |\xi_{1,R}(R_1)|^2 U(R_1-R_2) |\xi_{2,R}(R_2)|^2,   \label{U2}, \\
U_3&=&\int d^3R_1 d^3R_2 |\xi_{1,L}(R_1)|^2 U(R_1-R_2) |\xi_{2,R}(R_2)|^2. \label{U3},
\end{eqnarray}
Because we have only two states in each double-well, we can map the Hamiltonian onto one of interacting spin-1/2 particles using $\mathbf{\hat{S}}_i  \equiv  \left( \hat{a}^{\dagger}_{i,L}, \hat{a}^{\dagger}_{i,R} \right) \mathbf{\hat{\sigma}}  \left( \hat{a}_{i,L}, \hat{a}_{i,R} \right) ^T$ where $\mathbf{\hat{\sigma}} $ is the Pauli matrix given by
\begin{eqnarray}
\sigma^x=\begin{pmatrix}
0 & 1 \\ 
1 & 0 
\end{pmatrix},  \hspace{0.2cm}
\sigma^y=\begin{pmatrix}
0 & -i \\ 
i & 0 
\end{pmatrix}, \hspace{0.2cm}
\sigma^z=\begin{pmatrix}
1 & 0 \\ 
0 &-1
\end{pmatrix}.
\end{eqnarray}
As such, we see that $\hat{S}^x_i=\hat{a}^{\dagger}_{i,L} \hat{a}_{i,R} + \hat{a}^{\dagger}_{i,R} \hat{a}_{i,L}$ and $\hat{S}^z_i=\hat{n}_{i,L}-\hat{n}_{i,R}$. Since we have only one heavy atom in each double-well, i.e., $\hat{n}_{i,L}+\hat{n}_{i,R}=1$, we get $\hat{n}_{i,L}=(1+\hat{S}^z_i)/2$ and $\hat{n}_{i,R}=(1-\hat{S}^z_i)/2$. Substituting the above results to the extended Hubbard model, we get,
\begin{equation}
\hat{H}=\sum_{i=1}^{N_h} -\hbar \omega_0 \hat{S}^x_i+U_s \hat{S}^z_i \hat{S}^z_{i+1}+ \mbox{ const} \label{spinH}
\end{equation}
where $U_s=(2 U_2-U_1-U_3)/4$ and const $= (U_1+2U_2+U_3)/4$, which is the Hamiltonian of the quantum Ising model in a purely transverse field $-\hbar \omega_0$  \cite{Ising} and $U_s$ the effective nearest-neighbour interaction \cite{Greiner}. By shifting the relative positions of the two sublattices, it is easy to create an energy bias between left and right wells as in \cite{Bloch} so that we can also introduce the longitudinal field to the above model which makes this a versatile simulator for the quantum Ising model.

However, it is not our goal to study this model in detail here since there is already an abundant literature on this subject \cite{Ising}. Rather, it is presented here as an example of what can be studied with this electron-phonon simulator. \\

\subsection{ Interaction parameters of the effective models}

Up to this point, we have not discussed how to calculate the effective parameters $\omega_0$ and $U_s$. While $\omega_0$ can be controlled by $V_0$ and $V_1$, the effective interaction $U_s$ can be controlled by the filling of the light atoms to the energy band as shown below, which leads to interesting phases for the heavy atoms. Here, we just consider the full- and half-filling cases, which allow us to use a dimerized lattice to calculate the effective interaction. For other filling factors, e.g., 1/3 filling, a trimerized lattice in principle is needed to calculate the effective interaction and for even small filling factors, the approach would become very complicated.

In order to calculate the interaction between the heavy atoms in both the full-filled and half-filled band cases, we study the band structure of the dimerized lattice (see Fig.\ref{spinpeierls}).  Since each unit cell now has two atoms, the $c_j$ of Eq. (\ref{wf})  will be replaced by $c_j^{1, 2}$ where $1$ and $2$ refer to the two heavy atoms in the unit cell. In terms of the $1$ and $2$ atoms in each unit cell, Eq. (\ref{wf}) now reduces to 

\begin{gather}
\left( \kappa-\frac{1}{a} \right) c_0^{1}=c_0^{2} \frac{e^{-\kappa r}}{r}+  \sum_{j=1}^{\infty} c_j^{1}\frac{{\rm e}^{-\kappa (2jd) } }{2jd} +c_{-j}^{1}\frac{{\rm e}^{-\kappa (2jd) } }{2jd} \nonumber \\+ c_j^{2}\frac{e^{-\kappa(2jd+r)}}{2jd+r}+c_{-j}^{2}\frac{e^{-\kappa(2jd-r)}}{2jd-r}    \\ 
\left( \kappa-\frac{1}{a} \right) c_0^{2}=c_0^{1} \frac{e^{-\kappa r}}{r}+  \sum_{j=1}^{\infty} c_j^{2}\frac{{\rm e}^{-\kappa (2jd) } }{2jd} +c_{-j}^{2}\frac{{\rm e}^{-\kappa (2jd) } }{2jd}  \nonumber \\+ c_j^{1}\frac{e^{-\kappa(2jd-r)}}{2jd-r}+c_{-j}^{1}\frac{e^{-\kappa(2jd+r)}}{2jd+r} 
\end{gather}
where $r$ is the separation between the two heavy atoms in the unit cell. Using Bloch's theorem, $c_j^{1,2}=e^{ik 2jd} \eta^{1,2}$, the above equation becomes (in matrix form)

\begin{widetext}
\begin{gather}
\begin{pmatrix}
\frac{1}{a}-\kappa +\sum_{j=1}^{\infty} \frac{{\rm e}^{-\kappa (2jd) } }{2jd} \cos{2kjd} & \frac{e^{-\kappa r}}{r} +\sum_{j=1}^{\infty} (\frac{e^{-\kappa(2jd+r)}}{2jd+r}e^{ik 2jd}+\frac{e^{-\kappa(2jd-r)}}{2jd-r}e^{-ik 2jd})\\
\frac{e^{-\kappa r}}{r} +\sum_{j=1}^{\infty} (\frac{e^{-\kappa(2jd+r)}}{2jd+r}e^{-ik 2jd}+\frac{e^{-\kappa(2jd-r)}}{2jd-r}e^{ik 2jd}) &\frac{1}{a}-\kappa +\sum_{j=1}^{\infty} \frac{{\rm e}^{-\kappa (2jd) } }{2jd} \cos{2kjd} 
\end{pmatrix}
\begin{pmatrix}
\eta^{1}\\ \eta^{2}
\end{pmatrix}
=0
\end{gather}
\end{widetext}
Calling the matrix $\Xi(\kappa)$, the energy band structure is given by the condition of $\det [\Xi (\kappa)]$=$0$. 
When $r=d$, $\Xi_{12}$ and $\Xi_{21}^{*}$ are equal and reduce to 
\begin{equation}
e^{-ik}\sum_{j=1}^{\infty}\frac{e^{-\kappa(2j-1)d}}{(2j-1)d}\cos{[(2j-1)kd]}
\end{equation} so the equation determining the band structure becomes 
\begin{eqnarray}
\frac{1}{a}-\kappa+\sum_{j=1}^{\infty} \frac{{\rm e}^{-\kappa (2jd) } }{2jd} \cos{2kjd} = \nonumber \\
\pm \sum_{j=1}^{\infty}\frac{e^{-\kappa(2j-1)d}}{(2j-1)d}\cos{[(2j-1)kd]}
\label{bcal}
\end{eqnarray}
\noindent
the equation with minus sign on the right hand side of Eq. (\ref{bcal}) reduces to 
\begin{gather}
\frac{1}{a}-\kappa+\sum_{j=1}^{\infty} \frac{{\rm e}^{-\kappa jd } }{jd} \cos{kjd} \nonumber \\=\frac{1}{a}-\kappa-\frac{1}{d}\log{[1-2e^{-\kappa d} \cos{k}+e^{-2\kappa d}]}=0
\label{bcal2}
\end{gather}
where $\log(1-x)=-\sum_1^{\infty} x^n/n$ is used. The above equation can be recast to a quadratic equation in terms of $e^{-\kappa d}$, the solution of which is

\begin{equation}
\kappa=-\frac{1}{d}{\rm arccosh} [e^{d/a}/2+\cos(k)]
\end{equation}
so the band structure of Eq. (\ref{band})  is recovered. For the equation with plus sign in the right hand side of Eq. (\ref{bcal}), note that, by setting $k \rightarrow k+\pi$, it reduces to the same form of Eq. (\ref{bcal2}). So we get the same band structure and but shifted by $\pi$ in the Brillouin zone (see Fig.\ref{spinpeierls} b).

\begin{figure}[!hbp]
\centering
\includegraphics[width=\columnwidth]{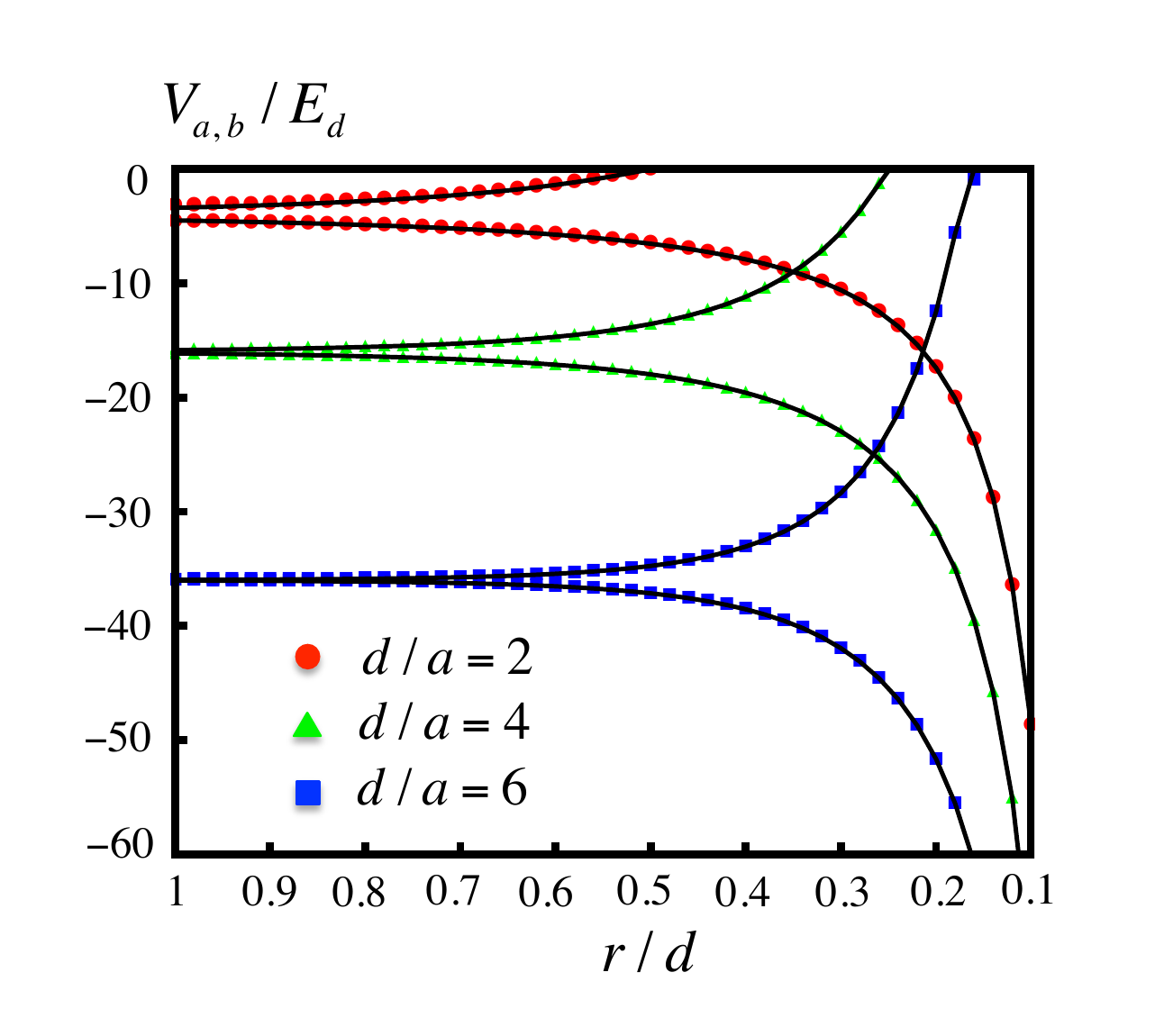}
\caption{ (Color online) The interaction of the heavy atoms, $V_a(r)$ and $V_b(r)$, in the dimerized lattice  fitted by the two-body potential, $-\hbar^2\kappa_-^2/2m$ and $-\hbar^2\kappa_+^2/2m$, for various values of $d/a$. Data points (circles, triangles and squares) are numerics from the integration over the band structure while the solid lines are the theoretical two-body potential of $-\hbar^2\kappa_-^2/2m$ and $-\hbar^2\kappa_+^2/2m$.}
\label{sup}
\end{figure}

When $r<d$, the single band for $r=d$ is split into a higher (a) and lower (b) bands. The analytical expressions of the band structure cannot be found in this case so we resort to numerics. The total interaction energy between heavy atoms is $V_{{\rm b}}(r)$ for a half-filled band and $V_{{\rm b}}(r)+V_{{\rm a}}(r)$ for full-filled band where $V_{{\rm a(b)}}(r)$=$\sum_{k} E_{{\rm a(b)}}(k, r)$ and $k \in [-\pi/2d,  \pi/2d]$.  We found empirically that $V_b$ and $V_a$ can be perfectly fitted by the symmetric $(\kappa_+)$ and antisymmetric $(\kappa_-)$ solutions of $\kappa_{\pm} \mp e^{-\kappa_{\pm} r}/r $=$1/a$ \cite{PetrovReview} (see Fig. \ref{sup} ). This means that the effective interaction in Eqs (\ref{U1},\ref{U2},\ref{U3}) can be given by
\begin{equation}
U(R)=V_a(R)+V_b(R)=-\frac{\hbar^2\kappa_-^2(R)}{2m}-\frac{\hbar^2\kappa_+^2(R)}{2m}
\end{equation}
for the full-filled band (which is nothing but Eq. (\ref{Petrov}), whereas in the half-filled case it is
\begin{equation}
U(R)=V_b(R)=-\frac{\hbar^2\kappa_+^2(R)}{2m}
\end{equation}
 So the attractive interaction in the half-filling case that leads to the dimerization studied in the following section has an interesting connection with the Efimov physics, since the dimerized Peierls phase can be viewed as a collection of trimers bound by the usual Efimov $1/R^2$ potential.

\subsection{FM and AFM phases at full and half-filling} 

After the discussion of the effective interactions of the models for the heavy atoms, we will now discuss several possible phases of the heavy atoms at full- and half-filling. When $\hbar \omega_0  \gg  U_s$, the spins point along $x$ and we recover the situation of the simple OL treated previously where the phonon frequency ($\sim  \omega_0$) is only weakly affected by interactions and the electron-phonon coupling strength, $(\omega^2(q$=$\pi/d)-\omega_0^2)/\omega_0^2  \simeq  (U_s/\hbar \omega_0)^2$ is very small.

In the opposite limit $\hbar \omega_0  \ll  U_s$ we can distinguish two cases: when the light atom band is full and when the band is half-filled leading to a dimerization of the lattice due to the  Peierls instability. 

In the full-filled  band case, $U_s$ is dominated by $-U_1$ and is negative. For $d/a=2$ and the previous superlattice parameters, we get $U_s \simeq  -0.6 E_d$. This corresponds to the ferromagnetic (FM) phase where all the spins point in the same direction along $z$ and oscillate slowly between the positive and negative directions with frequency $\omega_0$, becoming a symmetry-broken state in the limit $\omega_0 \rightarrow 0$. In the half-filled case $U_s>0$ which leads to an antiferromagnetic (AFM) state where neighbouring spins are anti-aligned along the $z$ direction. In terms of heavy atoms, it corresponds to a CDW, a dimerization of the lattice (Fig. \ref{spinpeierls}a). As is familiar from the Su-Schrieffer-Heeger (SSH) model, in this phase the light atoms are gapped which, for our parameters, is measurable $\simeq 4 E_d $ (Fig. \ref{spinpeierls}b). By shifting the relative positions of the two sublattices, it is easy to create an energy bias between left and right wells as in \cite{Bloch}. The validity of the BOA in the double-well case can be checked  in both full and half-filling cases. We must see whether the true gap ($E_{\rm gap}+ \mbox{ deformation energy of the lattice}$) is nonzero. In the full-filled case we compare the gain in energy of the lattice deformation: $-\hbar^2 \exp(-d/a)/2 m (0.6 d)^2 $ to $E_{\rm gap}$ from Eq. (\ref{gap}). It turns out that this energy is smaller by a factor 3.3 so that the true gap is nonzero and comparable to $E_{\rm gap}$. In the half-filled case however the lattice is already deformed and no energy can be gained by removing a light atom: any such excitation is forbidden by energy conservation so that the true gap is the Peierls gap. It is clear that, in the limit of large displacement, the many-body state reduces to a collection of trimers where the heavy atoms are attracted by the usual $1/R^2$ Efimov potential.

While the ``ion" motion uses a Hubbard-type (lattice model) description, however, we can go far beyond this since the light atoms are dynamical degrees of freedom beyond the BOA and which are not describable in terms of lattice particles. For example, we could study the dynamics of the Peierls instability itself and the time-dependence of the formation of the associated CDW, since this involves exciting some light atoms to conserve total energy. The kinetics of processes beyond the BOA are proportional to the matrix element squared between different light atom configurations, i.e., proportional to $(m/M)^2$. For typical values of mass ratio this is still observable over the lifetime of an experiment while creating a clear separation of timescales which allows us to study phenomena such as ``electron"-phonon scattering and transport, mobility of electrons and holes in the presence of umklapp processes.

\begin{figure}[!hbp]
\centering
\includegraphics[width=1\columnwidth]{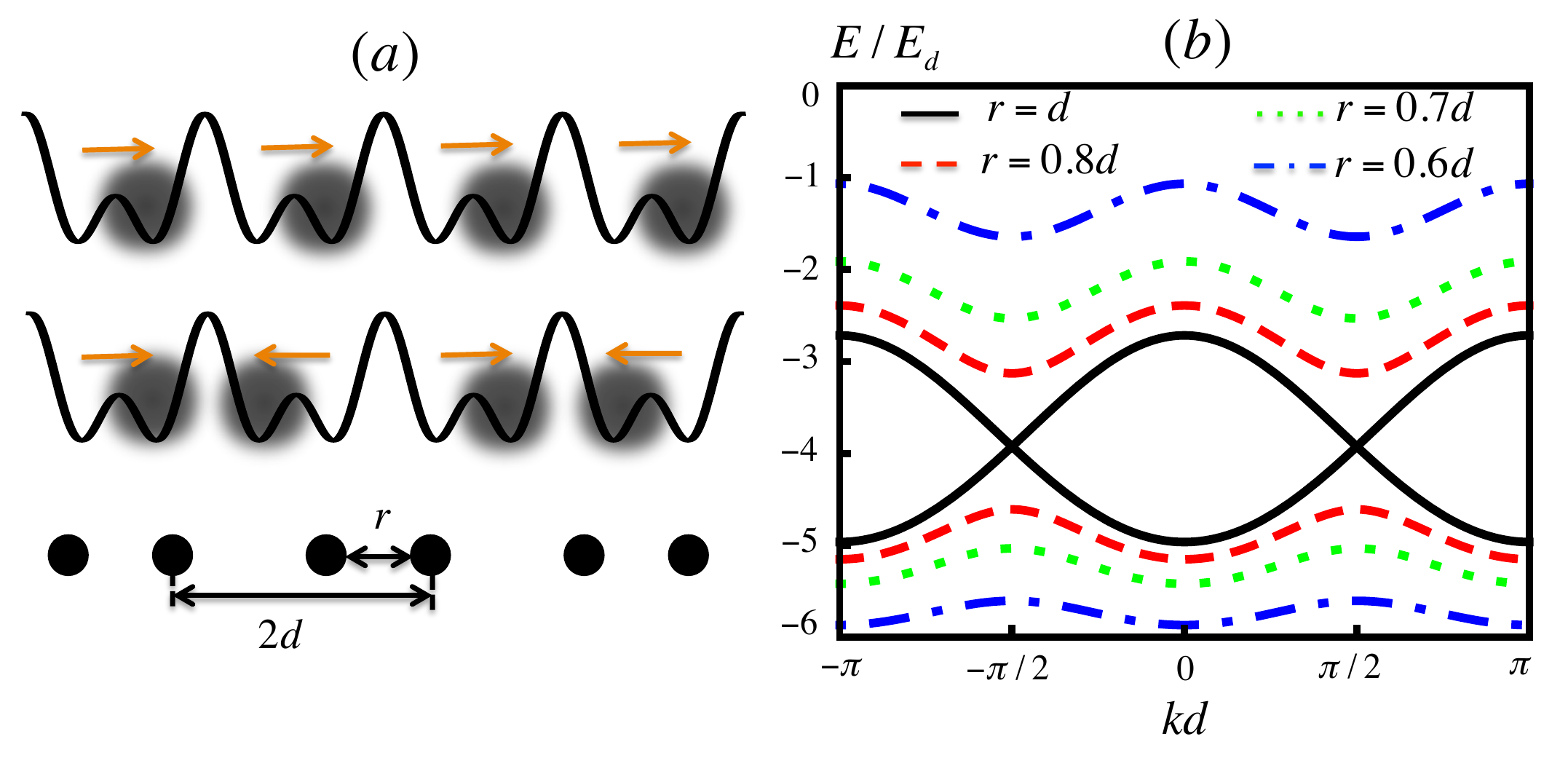}
\caption{ (Color online) (a) The FM/AFM phase of the heavy atoms at full/half-filling, and the dimerized lattice to calculate the interactions between the heavy atoms at these fillings. (b) Band structure of the dimerized lattice at $d/a=2$ for various values of $r$. The Peierls gap at $\delta \langle z \rangle$ $\simeq  0.2 d$, i.e, $r=0.6d$, is measurable $\simeq  4 E_d $.}
\label{spinpeierls}
\end{figure}

\section{Experimental issues}
\label{s5}
To prepare the system experimentally, we consider a 2D array of 1D tubes with a superlattice along each tube. Experimentally, a species-selective 1D OL has already been used to confine only one atomic species in 2D while having a negligible effect on the other species, which remains in 3D \cite{florence1,florence2}. Then we can either create the molecules with small $a$ in the gas phase and then load them into a deep OL so that we get an occupancy of one molecule per site or to first load the OL with heavy atoms, place them in contact with a gas of light atoms which then can, through three-body collisions, form Feshbach molecules on each site \cite{note7}. Our setup requires achieving scattering length on the order of lattice spacing $a \sim d$, for which we need a Feshbach resonance with good magnetic field control. A very large scattering length ($a \sim 600$ nm which is on the order of typical lattice spacings) has already been demonstrated even for a narrow resonance of width $\sim$1 G in a $^6$Li-$^{40}$K mixture \cite{rudi}. Other mass-imbalance Fermi-Fermi and Fermi-Boson mixtures, e.g., $^6$Li-$^{173}$K and $^6$Li-$^{174}$Yb \cite{LiYb1, LiYb2}, have also been realized experimentally. To prepare the half-filled band we could start with the full-filled band and then adiabatically increase $a/d$ so that the gap closes, lose half of the light atoms and then decrease $a/d$ to return to the gapped case. To measure the phonon dislocations we propose to scatter light off the heavy atoms. This will reveal of course the periodicity of the OL but a periodic deformation of the lattice (e.g., a CDW) would be visible as a second peak at the corresponding wave vector \cite{scattering1, scattering2}.
To measure light atom properties, we could use rf-spectroscopy \cite{rudi} which would reveal the energy distribution of the band and the Peierls gap.

\section{conclusions}
\label{s6}
We have shown how using ideas from few-body physics can bring a new approach to atoms in OLs, which we exemplified by creating a lattice for the light atoms via interactions with heavy atoms and by implementing an analog of an electron-phonon system using a superlattice configuration. In future we plan to study the problem of the formation of light atom Cooper pairs due to phonon exchange and the SSH model \cite{Janne}; the propagation of a single light atom along the chain of heavy atoms (which implements the polaron problem as an analog of its original lattice-electron system \cite{polaron}) and
other few-body systems focussing on the situation where the heavy atoms are allowed to tunnel to neighbouring sites, a case without parallel in solid state systems where the ions are fixed to the lattice sites.

\begin{acknowledgements}  We would like to thank J. Dalibard, G. Ferrari, R. Hulet, V. B. Shenoy  and D. Jaksch for discussions. We acknowledge support from the EPSRC through grant EP/I018514/1.
\end{acknowledgements}


\begin{references}
\bibitem{note1} Note that the ``electron" and ``ion" energy scales, $E_d$ and $E_R$, are of the same order for $^6$Li- $^{40}$K, $^{87}$Rb mixtures but, for $^6$Li-$^{173}$Yb the ``electron" energy begins to dominate.
\bibitem{note2} The opposite limit $\omega_\perp \ll \omega_0$ would lead to transverse phonons instead of the longitudinal ones discussed here.
\bibitem{note3} This requirement is perhaps too strict and we plan to investigate the situation where the heavy atoms are allowed to tunnel.
\bibitem{note4} When $a$ is comparable to the van der Waals radius, the polarizability of the heavy atom (and hence its trapping potential) changes due to molecular effects, so we consider only larger values of $a$.
\bibitem{note5} Peierls' instability prevents the closing of the gap due to the coupling to the heavy atoms. However this coupling is very weak so that the actual gap will be small.
\bibitem{note6} Curiously, there can be a dissociation due to gravity of the heavy-light dimer (or, equivalently, the tunnelling of light atoms out of the band) which occurs when $a^3 \gtrsim \hbar^2/m^2g$. For $^6$Li this is around $50 \mu $m but for $^{173}$Yb it drops to $5 \mu$m, not so far from the currently achievable scattering lengths.
\bibitem{note7} Even though the light atoms are spin-polarized fermions, there can be an appreciable dimer formation rate in light-light-heavy collisions if the masses are different. Because we assume that $a$ is large, for the light atom gas we can easily enter the regime $na^3 \sim 1$, thereby increasing the rate appreciably.
\end{references}
\end{document}